\documentstyle[aps,psfig]{revtex}

\begin{document} 
\title{\Large Pair correlations in high-K bands} 
\author{\sc D.~Almehed, S.~Frauendorf and F.~D\"{o}nau} 
\address{\sl IKH, FZ Rossendorf, PF 510119, D-01314 Dresden, Germany 
and Department of Physics, University of Notre Dame, IN 46556, USA} 
 
\maketitle 
 
\begin{abstract} 
The tilted axis cranking model is used in combination with the random 
phase approximation and particle number projection to analyze the 
influence of dynamical pair correlations in the high-$K$ bands of 
$^{178}W$ and their effect on relative energy and angular 
momentum. The calculations show the importance of dynamical pair  
correlations to describe the experiment as well as advantages and 
problems with the different models in the superfluid and normal state 
regions.  
\end{abstract} 
 
 
\section{INTRODUCTION} 
The transition of a nucleus from the superfluid to the normal state at 
high angular momentum is an interesting problem that is studied by 
means of modern $\gamma$-detector arrays. Contrary to the analogous 
transition in solids, in the finite nuclear system there is no sharp 
phase change but an extended transition region with in which  
pairing effects 
disappear. The most rapid attenuation of pair correlations is caused by 
quasiparticle excitations i.e. breaking of pairs. This case is 
realized in deformed nuclei when  
a large fraction of the angular momentum is generated along the 
symmetry axis of the nucleus. These states may appear as  
high-$K$ isomers near the yrast line. Hence, the experimental data on 
high-$K$ isomers and the rotational structures built on those states 
contain valuable information about the pair correlations and how they 
are influenced by the rotation.  
 
The theoretical analysis of high-$K$ band structures presented in this 
paper is based on the  tilted axis cranking~\cite{Fr93} (TAC) model, 
which is a mean field approach for describing both the 
rotation and pair correlations in the framework of the 
Hartree-Fock-Bogoliubov (HFB) theory. In fact, the TAC model has 
already been applied to the high-$K$ multi-quasiparticle bands in 
$^{178,179}W$~\cite{FN00}. There, only the effects of {\it static} pair 
field have been considered. The central aim of our present 
investigation is  studying the role of {\it dynamical} pair 
correlations. For this purpose we
apply the random phase approximation (RPA) to the pairing 
interaction, which  includes the fluctuations of the pair field. 
The fluctuations are particularly important when the static 
pair field has collapsed. Their relevance is 
suggested by the results of \cite{FN00,SG89} for  $^{178,179}W$ 
as well as in the earlier investigations of the pair correlations of 
other high-$K$ band heads states~\cite{JB95}. 
 
The combination of 
the HFB theory with RPA does not provide an reliable  
description in the region where the  
pairing gap disappears ~\cite{RS80}. There 
  the  particle number 
projection method (PNP) works  better ~\cite{SG89,BD86}. Therefore, both 
methods, RPA and PNP, are considered and compared with each other.  
 
The paper is organized as follows.  
In section~\ref{model} we develop TAC versions which include 
pairing-RPA or PNP. Details of the calculations are given in section 
~\ref{parameters}.  
The results of the 
calculations of high-$K$ bands in $^{178}W$ are presented in 
section~\ref{calc} and compared with the experiment~\cite{PW98}.  
 
\section{THE MODEL} 
\label{model} 
\subsection{Mean field} 
Our investigations are based on the  
single particle Routhian~\cite{Fr93} of the TAC model complemented by
a monopole pair interaction term, 
\begin{equation} 
        {\cal H}' = t + V(\varepsilon_\nu) - \omega(\sin \vartheta j_1 + \cos 
        \vartheta j_3) - G P^{\dagger}P - \lambda \hat{N},  
        \label{full-routhian} 
\end{equation} 
where  $P^{\dagger}$ denotes the monopole pair field 
operator~\cite{RS80} and $\hat{N}$ is the particle number operator.
For simplicity the terms are written only for one kind of particles. 
 
Replacing in ${\cal H'}$ the pairing two-body term by the pair 
potential one obtains the  quasiparticle Routhian  
\begin{equation} 
        h' = t + V(\varepsilon_\nu) - \omega(\sin \vartheta j_1 + \cos \vartheta 
        j_3) - \Delta(P^{\dagger}+P) - \lambda \hat{N}.  
        \label{qp-routhian} 
\end{equation} 
The  diagonalization of $h'$ (for details cf.\cite{RS80}) 
provides the quasiparticle energies 
$e'_i(\omega)$ as well as the HFB amplitudes ($u,v$),  
required later on for the RPA 
and PNP calculations. In order to define a  
quasiparticle state $|>$ for a specific configuration    
the occupation numbers need to be chosen.  
The self-consistent treatment of $h'$ implies determining    
the pair gap $\Delta$ from the HFB gap equation 
\begin{equation} 
        G<P^{\dagger}> = \Delta 
        \label{scDelta} 
\end{equation}
and fixing the chemical potential $\lambda$ by the particle number  
condition   
\begin{equation} 
        <\hat{N}> = N. 
        \label{scN} 
\end{equation} 
These conditions remain valid for the RPA but they will be modified 
for the PNP approach (cf. section \ref{PNPmodel}). The 
HFB part of the total Routhian becomes  
\begin{equation} 
        R_{HFB} = <{\cal H'}> + \lambda N. 
        \label{scRTAC} 
\end{equation} 
In order to determine the deformation parameters $\varepsilon_\nu$ 
we have constructed the Total Routhian Surface (TRS) using  
the standard Strutinsky renormalization procedure as described e.g. 
in \cite{NP76,Sz83}.    
As shown in~\cite{WD95} the essential shell-correction part  
$E_{strut}(\varepsilon_\nu)$    of the TRS  
can be calculated for a non-rotating and unpaired 
ground state. The contribution containing the dependence on rotation and  
pair field enters via the quasiparticle Routhian, eq.~(\ref{scRTAC}).  
Hence, the total Routhian can be written as        
\begin{equation} 
        R(\omega,\varepsilon_\nu,\Delta,\lambda,\vartheta) =  
E_{strut}(\varepsilon_\nu) + 
        R_{HFB}(\omega,\varepsilon_\nu,\Delta,\lambda,\vartheta). 
        \label{totalR} 
\end{equation} 
   
\subsection{Pairing-RPA} 
\label{RPAmodel} 
\begin{figure}[htb] 
     \centerline{\psfig{figure=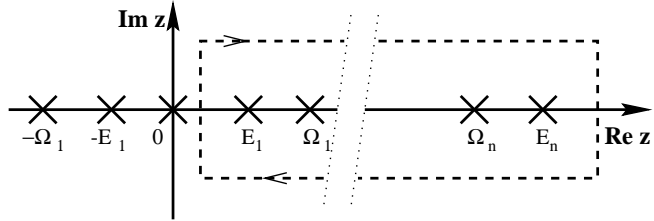,height=3cm}} 
  \caption{A schematic picture of the integration contour (dashed 
    line) in the complex plane. The roots $\Omega_\nu$ and poles 
    $E_\mu$ of $F(z)$ are marked with crosses. }  
    \label{fig:int-cont} 
\end{figure}

So far only the mean field part of the pairing 
energy has been taken into account. Now we include the pair 
correlation energy, which comes from quantum vibrations 
around the mean field minimum.  
The RPA treatment of the pair interaction gives the following 
expression for the correlation energy~\cite{SG89}  
\begin{equation} 
  \label{eq:rpa-corr} 
  E_{corr}^{RPA} = \frac{1}{2} \left[ \sum_\nu \Omega_\nu - \sum_{\mu} 
    E_\mu \right] + E_{ex}  
\end{equation} 
where $\Omega_\nu$ are the RPA frequencies for the  pair vibrations, 
$E_\mu=e_i+e_j$ ($\mu=i<j$) are the two-quasiparticle energies and  
$E_{ex}$ is the so-called boson exchange term. The total Routhian 
is the obtained from eqs.~(\ref{totalR},\ref{eq:rpa-corr}) as  
\begin{equation} 
  \label{eq:Erpa} 
  R^{RPA}(\omega,\varepsilon_\nu,\Delta,\lambda,\vartheta) = 
  R(\omega,\varepsilon_\nu,\Delta,\lambda,\vartheta) + 
  E_{corr}^{RPA}(\omega,\varepsilon_\nu,\Delta,\lambda,\vartheta).  
\end{equation} 
Due to the large number of RPA roots in deformed nuclei, it is practically  
impossible to evaluate the 
sum in eq.~(\ref{eq:rpa-corr}) directly \cite{SG89,BD86}. However, it
 can be calculated by means of the integration method developed  
recently~\cite{DA99}, 
which is especially simple for pairing interaction. It uses the RPA 
response function $F(\Omega)$ (cf.\cite{SG89,RS80}), which reads 
explicitly for our case: 
\begin{equation} 
  \label{eq:disp} 
  F(\Omega) = \left(2G \sum_\mu 
    \frac{S_\mu^{+^2}E_\mu}{E_\mu^2-\Omega^2}-1\right)\left(2G\sum_\mu 
    \frac{S_\mu^{-^2}E_\mu}{E_\mu^2-\Omega^2}-1\right) - \left(2 G 
    \Omega \sum_\mu \frac{S_\mu^+S_\mu^-}{E_\mu^2-\Omega^2} \right)^2  
\end{equation} 
where 
\begin{equation} 
  \label{eq:spm} 
  S_\mu^\pm = \frac{1}{\sqrt{2}} \sum_k \left\{ \left( u_{ki} 
      u_{\bar{k}j} \pm v_{kj} v_{\bar{k}i} \right) - \left( u_{kj} 
      u_{\bar{k}i} \pm v_{ki} v_{\bar{k}j} \right) \right\}   
\end{equation} 
and $u,v$ are the HFB amplitudes. The zeros of the function 
$F(\Omega_\nu)$ determine the the RPA frequencies, $\Omega_\nu$.  
Continuing the variable $\Omega$ into the complex plane $z$,  
one defines the spectral function $\frac{F'(z)}{F(z)}$.  
It has first order poles at $\Omega_\nu$ and $E_\mu$ and it is 
analytical for all other complex values of $z$. According to  
Cauchy's theorem the following integral relation is obtained: 
\begin{equation} 
  \label{eq:g_int} 
  \frac{1}{2\pi i} \oint_C \!\!\! dz \,\,g(z) \frac{F'(z)}{F(z)} = 
  \sum_\nu g(\Omega_\nu) - \sum_\mu g(E_\mu) , 
\end{equation} 
where $g(z)$ is a arbitrary complex function which is analytical 
within the region enclosed by the integration path $C$. The roots 
$\Omega_\nu$ and the poles $E_\mu$ of $F(z)$ lie in the same region. By 
choosing $g(z)=z$ the RPA correlation energy, eq.~(\ref{eq:rpa-corr}) 
becomes   
\begin{equation} 
  \label{eq:rpa-int} 
  E_{corr}^{RPA} = \frac{1}{4\pi i} \oint_C \!\!\! dz z 
  \frac{F'(z)}{F(z)} + E_{ex} , 
\end{equation} 
where the integration path $C$ goes around the right half of the 
complex plane. The exchange term is given by  
\begin{equation} 
  \label{eq:Eex} 
  E_{ex} = \frac{G}{2} \sum_\mu \left( S_\mu^{+^2} + S_\mu^{-^2}\,, 
  \right) 
\end{equation} 
as in~\cite{SG89}. The integral in eq.~(\ref{eq:g_int}) is independent 
of the path $C$ as long as all poles in the positive plane are 
enclosed (see Fig.~\ref{fig:int-cont}). Therefore $C$ can always be 
chosen in such a way that the spectral function becomes smooth and the 
integration numerically stable, such that a small number of grid points 
can be used in the integration.  
 
The RPA correction to the angular momentum operator ~\cite{Na87} 
\begin{equation} 
  \label{eq:JRPA} 
  J_i^{(2)} = \sum_{ij, kml} j_{km}^i \left( u_{ki} u_{mj} - v_{ki} 
    v_{mj} \right) b^\dagger_{il} b_{jl}, 
\end{equation}  
 gives us the RPA contribution 
 $<J_i^{(2)}(\Omega_\nu)>$  of the RPA solution $\Omega_\nu$, 
where $j_{km}^i$ are the single particle matrix elements of the angular 
momentum components $i=x,y,z$ and the term $b^\dagger_{il}b_{jl}$ is 
the boson image of the two quasiparticle scattering contribution of 
the angular momentum operator. 
By choosing the weight function $g(z)= <J_i^{(2)}(z)>$ in the integral,   
 (\ref{eq:g_int})  
the corresponding RPA contribution to $J_i$ can be calculated too.  
 
It should be pointed out that eq.~(\ref{eq:rpa-corr}) takes into account 
also the contribution from possible spurious modes. In the paired case  
$\Delta >0$ the function $F(z)$ has a root at $z=0$. This does not 
cause a problem for the integration since the factor $z$ in the 
integrand in eq.~(\ref{eq:rpa-int}) cancels this pole. After the 
transition to $\Delta =0$ the spurious mode disappears.

The above described  method can be straightforwardly applied to obtain  
the correlation energy of the $K=0$ ground state band. 
The RPA for the excited configurations of the high-$K$ bands 
needs special care concerning the selection of the ``physical'' 
phonon modes from the full RPA spectrum $\pm\Omega_\nu$ which implies 
positive and negative frequencies. The general criterion is to select
from the pair of roots  $\pm\Omega_\nu$  the one with the positive normalization.  
Such an identification is trivial for the ground configuration because 
there all $\Omega_\nu>0$ have positive normalization.  
However, for excited configurations the normalization check has to be 
done explicitly. Since the excited configurations under study are 
still close to the ground configuration the search for 
the possible physical phonons with $\Omega_\nu<0$ can be confined to a  
narrow energy region of a few MeV above zero. For this region the sum  
(\ref{eq:rpa-corr}) is directly evaluated whereas for the  
higher lying part the loop integration is done.   
We mention that the correlation energy is also for excited 
configurations a negative 
(gain) term the value of which depends on the rotational frequency as 
well as on the configuration under study. 
 
In calculating the RPA contribution to the angular momentum components 
by means of eq.~(\ref{eq:JRPA}), we observed that 
for some phonons, $\Omega_\nu$, the backward going amplitudes become 
so large that the underlying quasiboson approximation (QBA) of the RPA  
is no longer valid. This happens in excited configurations 
near instabilities of the static pair field, which occur when two
configurations with different $\Delta$ cross each other.
Examples are  the bands  $K=7,15$ (neutron pairing) and 
$K=15,22$ (proton pairing), which will be discussed in section \ref{71522}. 
For the low lying RPA solutions we
check explicitly  whether the QBA condition, 
$\left< \nu \right| b_{il}^\dagger b_{jl} \left| \nu \right> \ll 1$, 
is satisfied or not. We exclude roots with 
$\left< \nu \right| b_{il}^\dagger b_{jl} \left| \nu \right> > 0.1$
from the sum (\ref{eq:JRPA}). 

This procedure  may be justified as follows.
In the extended boson approximation (EBA)~\cite{Ha64,IU65} the non  
zero term  $\left< \nu \right| b_{il}^\dagger b_{jl} \left| \nu \right>$  
is approximatively taken into account  by an iterative process.  
The relevant effect of the EBA as compared  to QBA is a strong reduction  of the 
backward going amplitudes~\cite{IU65}. A full EBA treatment 
is too demanding for 
a realistic calculation like ours. Instead we use the rough 
approximation that the backward going amplitudes  
are quenched in the cases when QBA is not valid. i.e.  if
$\left< \nu \right| b_{il}^\dagger b_{jl} \left| \nu \right> > 0.1$
 the roots are removed from the  sum (\ref{eq:JRPA}) 
of the RPA correction.
These roots would give a very  
large (several units of $\hbar$) contribution to  
the angular momentum. If they are included, $J_{RPA}$ changes very rapidly
near the  crossings between configurations with different $\Delta$.
It is known the cranking model is unreliable near such crossings,
where it seems justified to resort  to an only rough correction of the QBA,
which results in a smooth reasonable function $J_{RPA}(\omega)$.

\subsection{Particle number projection} 
\label{PNPmodel} 
The above treatment of the dynamical pair correlations  
by combining the HFB and RPA methods does not work
for $\Delta \rightarrow 0$. Another approximate method to include the 
dynamical pair correlations is the particle number projection 
(PNP)~\cite{SG89,RS80}. In the PNP approach a variational state $|N>$ 
with good particle number $N$ is formed by applying the projection 
integral   
\begin{equation} 
        |N> \,\,\, \propto \int^{2\pi}_{0}\!\!\! d\phi\,\,e^{- i 
          \phi(\hat{N}-N)} |>  
        \label{projection} 
\end{equation} 
to a TAC eigenstate $|>$ of the quasiparticle Routhian $h'$, 
eq.~(\ref{qp-routhian}). Number projection accounts for gauge angle 
fluctuations of the pair field on a circle in the complex plane 
($\Delta e^{- i \phi}$) with the radius $\Delta$, but disregards radial 
fluctuations.    
 
The PNP energy function to be minimized is the expectation value   
\begin{equation} 
        R_{PNP}(\Delta) = <N|{\cal H'}|N> + \lambda N 
        \label{pRTAC} 
\end{equation} 
calculated with the projected state $|N>$, eq.~(\ref{projection}).  
We   do not perform a full variation of the HFB amplitudes of 
the reference state $|>$ but minimize the energy 
(\ref{pRTAC}) only with respect to the variable gap parameter 
$\Delta$. The minimization of eq.~(\ref{pRTAC}) in PNP replaces the  
self-consistency condition (\ref{scDelta}) for calculating $\Delta$ 
in the HFB approach. Compared to the self-consistent pair gap (denoted 
by $\Delta^{HFB}$ in the following) of an unprojected  quasiparticle 
state, the pair gap $\Delta^{PNP}$ optimizing the projected Routhian 
(\ref{pRTAC}) is generally larger and stays non-zero through the phase 
transition region.  
 
\begin{figure}[hbt] 
           \centerline{\psfig{figure=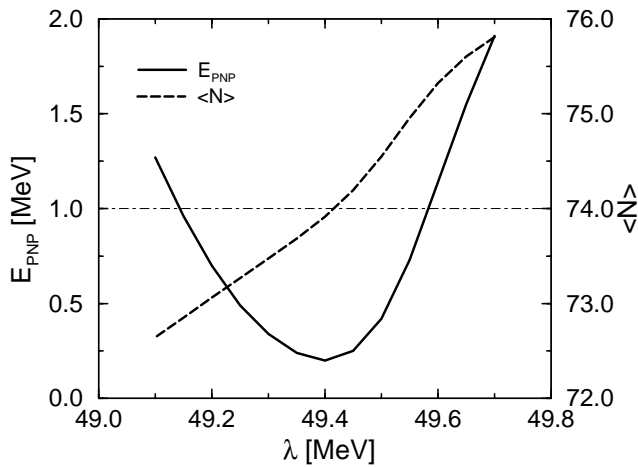,height=6.5cm,angle=-90}} 
         \caption{Particle number expectation value in the unprojected 
           case, $<N>$, and the total energy in the projected  case, 
           $E_{PNP}$, as functions of $\lambda$. This example 
           corresponds to the yrast neutron configuration at 
           $\omega=0.4$ (MeV). The correct particle number $<N>$ is 
           represented by a horizontal dot-dashed line.}  
         \label{E-lambda} 
\end{figure}

In the HFB variant of TAC, the constraint $<\hat N>=N$  
adds an additional dimension to the system of  
nonlinear equations given by eq.~(\ref{scDelta}) and the minimization 
with respect to $\varepsilon_\nu$ and $\vartheta$. Performing the  
PNP the above condition is automatically satisfied. Instead $\lambda$ 
becomes  another variable parameter of the Ritz variational problem,  
which is to be determined by minimizing the Routhian. However, the 
minimum practically coincides with the solution of eq.~(\ref{scN}), as 
illustrated in Fig.~\ref{E-lambda}. This has the advantage that in the 
vicinity of the minimum the energy does not change very much.   
Thus the  errors remain small if the minimum in $\lambda$ is not 
exactly found (or eq.~(\ref{scN}) is not exactly solved). Keeping 
$\lambda$ fixed greatly simplifies the actual calculation. However, 
one has to be careful that fixing $\lambda$ does not affect the 
general properties of the configuration too much. As will be discussed 
in Appendix~\ref{PNP}, neglecting the exchange term of the interaction  
may lead to problems in calculating the projected energy. Unlike the 
unprojected HFB function  
$|>$, the projected wave function~(\ref{projection}) is not be  
stationary. This may lead to problems that will be discussed   
in section~\ref{stationary}. 
 
\boldmath 
\subsection{Energy minimum in $\vartheta$} 
\unboldmath 
\label{stationary} 
Since the shell correction part in total Routhian $R$, eq.~(\ref{totalR}), does 
not depend on the tilt angle $\vartheta$ one has  
\begin{equation} 
        \frac{\partial R}{\partial \vartheta} = \frac{\partial}{\partial 
          \vartheta} \left< h' \right> = \left< \frac{\partial 
            h'}{\partial \vartheta} \right> + 2 <|h' 
        \frac{\partial}{\partial \vartheta}|> . 
        \label{dE'dtheta} 
\end{equation} 
The term $<| h'\frac{\partial}{\partial \vartheta} |>$ vanishes because 
$|>$ is a stationary eigenstate to $h'$. Hence, the derivative of the 
Routhian becomes  
\begin{equation} 
        \frac{\partial R}{\partial \vartheta} = \left< \frac{\partial 
            h'}{\partial \vartheta} \right> = -\omega \left( \cos\vartheta 
          <J_{1}> - \sin\vartheta <J_{3}> \right) = -\omega J_{\perp}.  
        \label{static} 
\end{equation} 
Thus, the requirement $\frac{\partial R}{\partial \vartheta} = 0$ for the 
minimization is cast into the condition $J_{\perp}=0$. In other words, 
$\vec{\omega}$ and $\vec{J}$ must be parallel~\cite{Fr93}.   
 
In case of particle number projection, the state $|N>$ is not an 
eigenstate to $h'$ and  
eq.~(\ref{static}) is not strictly valid. The energy minimum will no 
longer exactly agree with the condition of parallelity, 
which is the condition for uniform rotation. In the PNP calculations 
the minimum condition $\frac{\partial R}{\partial \vartheta} = 0$  
needs to be  
fulfilled even if $J_{\perp}=0$ is not satisfied.  
 
We generally found small differences ($0^\circ-10^\circ$) between the 
values of $\vartheta$ obtained from the condition 
$J_{\perp}(\omega)=0$ and the energy minimum. Substantial deviations 
appear in regions of band crossings, where the cranking model is 
unreliable~\cite{Ha76} anyway, and sometimes close to the band head.  
 
\section{DETAILS OF THE CALCULATIONS} 
\label{parameters} 
\begin{figure}[htb] 
           \centerline{\psfig{figure=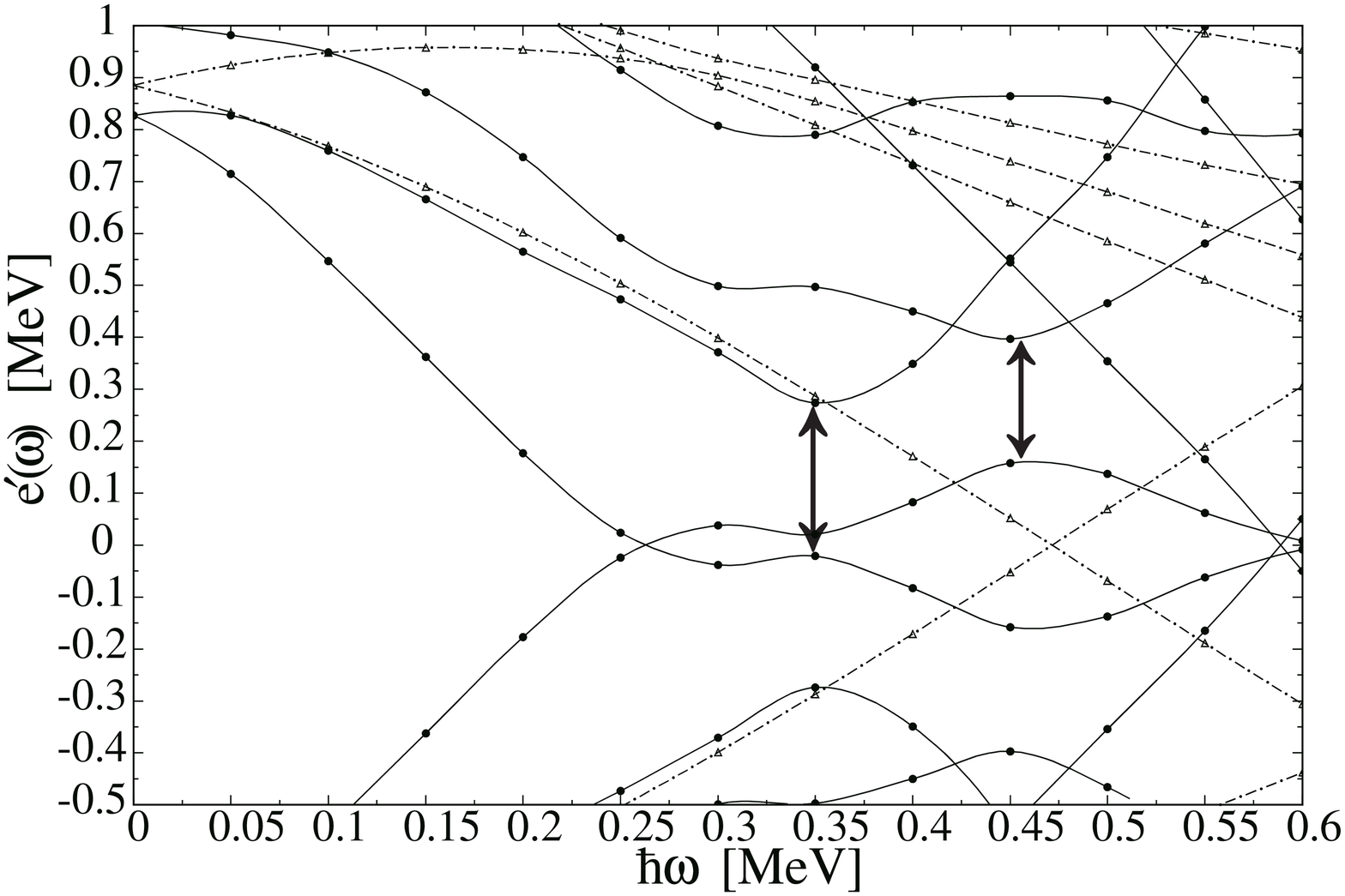,width=8cm}} 
         \caption{A quasiparticle diagram. The lines are connected by  
           diabatic tracing, i.e. by finding the largest overlap 
           between the points of two successive values of 
           $\omega$. Solid lines correspond to quasiparticles with 
           positive parity while the dot-dashed ones have a  
           negative parity. Two quasi crossings are marked with double 
           arrows.}  
        \label{spagetti1} 
\end{figure}

We use the modified oscillator model with deformations  $\varepsilon_2$, 
$\varepsilon_4$ and standard Nilsson parameters 
~\cite{NR95}. The high-$K$ bands in nucleus $^{178}$W are found to be 
axially symmetric  configurations. The tilt angle  
$\vartheta$ is the  angle between the symmetry (3-)axis 
and the rotational axis. 
The Strutinsky part $E_{strut}(\varepsilon_\nu)$ of the TRS, 
eq.~(\ref{totalR}) is  
obtained practically by including 8 oscillator shells for protons 
and neutrons, respectively. For calculating the quasiparticle  
term $R_{TAC}$, eq.~(\ref{scRTAC}),  
it is sufficient to consider only a few shells around the Fermi surface.  
We included in the diagonalization of the  
quasiparticle Routhian the four N-shells closest to the Fermi surface.  
It is important to follow a certain 
quasiparticle configuration when seeking the minimum of 
$R(\omega,\varepsilon_2\,\varepsilon_4,\Delta,\lambda,\vartheta)$. 
This is achieved by 'diabatic tracing': When changing one of the 
parameters determining $h'$ the overlap of the quasiparticle wave 
functions with  the  
ones before the step is calculated. By looking for the maximal overlap 
a one-to-one correspondence between the quasiparticle states is 
established. Figure~\ref{spagetti1} shows a quasiparticle diagram 
constructed in this way by using the step size $\Delta\omega = 
0.05$ MeV, $\Delta\vartheta = 5^{\circ}$. Keeping the occupation 
of such diabatic quasiparticle trajectories one usually follows a 
quasiparticle configuration of a given structure. Problems may appear 
near quasi-crossings like the ones in Fig.~\ref{spagetti1} where the 
configurations are mixed up.  
 
We have used two different strategies for finding the minimum with 
respect to $\Delta$ and $\vartheta$ for different $\omega$.  
\begin{enumerate} 
\item 
        \begin{enumerate}  
        \item Construct the configuration for $\Delta=0$. Choose a 
          reasonable start value $\vartheta_S$ for different $\omega$.   
        \item Find minimum in $\vartheta$ by tracing 
diabatically for fixed $\Delta$. 
        \item Find minimum in $\Delta$ by tracing diabatically 
for fixed $\vartheta$. 
        \item Iterate by returning to (b) until total minimum is found. 
        \end{enumerate} 
\item
        \begin{enumerate}  
        \item Construct the configuration for the band head for a 
          guessed $\Delta$. 
        \item Find minimum for  $\Delta$.
        \item Go to the next $\omega$-point with fix $\Delta$, tracing 
          diabatically.  
        \item Vary $\Delta$ and $\vartheta$  as in 1b - 1d.  
        \item Iterate by returning to (c)
        \end{enumerate} 
\end{enumerate} 
Usually 1. is better to work with because the minimum in $\vartheta$ does 
not  change   
 much when pairing is added.  Strategy 2. is more appropriate for the 
 ground configuration, because  
it has no  counterpart in the case of zero pairing. 
 
The equilibrium shape of $^{178}W$ turned out to be  
rather stable. The deformation differs
by less then 5$\%$  between different band heads (cf. \cite{XW98}). 
We  also found that the deformation changes 
very little within the bands. Thus, we  adopted the  fixed deformations
$\varepsilon_{2}=0.229$ and $\varepsilon_{4}=0.034$
calculated for the ground state. These values 
were used for all bands except  $K^{\pi}=25^{+}$. This band
contains a $h_{9/2}$ aligned proton,  
which drives the equilibrium values to $\varepsilon_{2}=0.255$ and 
$\varepsilon_{4}=0.038$. The relative energy of the bands is  sensitive to 
the strength of the pairing force $G$~\cite{XW98}.  In our calculations, $G$ 
was fixed to match the values of the even-odd mass differences
as calculated with PNP~\cite{Fr76}.   
\[\Delta_{\nu} = 1.15 \mbox{ MeV} \rightarrow  G_{\nu} = 0.093 \mbox{ 
  MeV}\]  
\[\Delta_{\pi} = 1.23 \mbox{ MeV} \rightarrow  G_{\pi} = 0.121 \mbox{ 
  MeV}\]  
 
In the PNP calculation the value of $\lambda$ is determined for each 
configuration at $\omega=0.4$ (MeV) by minimizing 
$R(\lambda)$ and kept constant for each configuration. In RPA and HFB 
calculations, $\lambda$ is adjusted according to eq.~(\ref{scN}) 
for each configuration and $\omega$, because the HFB energy is much more 
sensitive to this parameter.  
 
\section{RESULTS} 
\label{calc} 
The configurations of the considered $K$-bands are listed in 
table~\ref{config}. They are in accordance with~\cite{PW98}.
The following presentation is divided into two 
parts. The first part treats the cases where HFB gap $\Delta^{HFB}$ has 
collapsed already at the band head ($K^\pi=25^+$ and $30^+$) while 
second  comprises the cases where the pair field has transitional 
character ($K^\pi=7^-$, $15^+$ and $22^-$).  
 
For a given configuration, there is usually only a weak systematic reduction 
of $\Delta$ when 
$\omega$  increases.  It only changes substantially when a pair of protons or neutrons 
becomes aligned at a band crossing in a configuration with a low 
number of excited quasiparticles. For example, in the yrast band there 
is a drastic decrease of $\Delta_{\nu}$ through the crossing
region at $\omega_c \approx 0.3$. It is well known~\cite{Ha76} that the 
cranking model has problems  close to a 
band crossing. In this region, the  different strategies 
(cf. section~\ref{parameters}) of optimizing the mean-field parameters 
can result in different configurations and, as a consequence, in
different self-consistent values of the parameters.  
In our TAC calculations for $K$-bands, we observed the general tendency 
that the tilt angle $\vartheta$ increases monotonically  from 
$0^{\circ}$ at the band head to a value close to $\vartheta=90^{\circ}$ 
at highest frequency. In some cases there is a decrease in $\vartheta$ at 
high $\omega(>0.5)$, which can be related to the crossing with other 
configurations. 
\begin{table}[htb] 
\begin{center} 
\begin{tabular}{crl} 
$K^{\pi}$ & Neutron configuration       & Proton configuration\\ \hline 
$7^{-}$   & $\nu\{7/2^{+},7/2^{-}\}$ & \\ 
$15^{+}$  & $\nu\{7/2^{+},7/2^{-}\}$ & $\pi\{7/2^{+},9/2^{-}\}$ \\  
$22^{-}$  & $\nu\{5/2^{-},7/2^{+},7/2^{-},9/2^{+}\}$ & 
$\pi\{7/2^{+},9/2^{-}\}$ \\  
$25^{+}$  & $\nu\{5/2^{-},7/2^{+},7/2^{-},9/2^{+}\}$ & 
$\pi\{1/2^{-},5/2^{+},7/2^{+},9/2^{-}\}$ \\   
$30^+$    & $\nu\{5/2^{-},7/2^{+},7/2^{-},9/2^{+}\}$ & 
$\pi\{5/2^{+},7/2^{+},9/2^{-},11/2^{-}\}$ \\   
\end{tabular} 
\caption{Occupied orbitals $(\Omega^\pi)$ of the multi quasiparticle 
  excitations ($K^\pi$). The quasiparticle configurations are taken 
  from~\protect\cite{PW98}. Complete assignment:  
  Neutrons $5/2^{-}[512],7/2^{+}[633],7/2^{-}[514],9/2^{+}[624]$ and 
  protons 
  $1/2^{-}[541],5/2^{+}[402],7/2^{+}[404],9/2^{-}[514],11/2^-[505]$}  
\label{config} 
\end{center} 
\end{table}   
 
\boldmath 
\subsection{$K^\pi=25^+$ and $30^+$} 
\begin{figure}[hbt] 
           \centerline{\psfig{figure=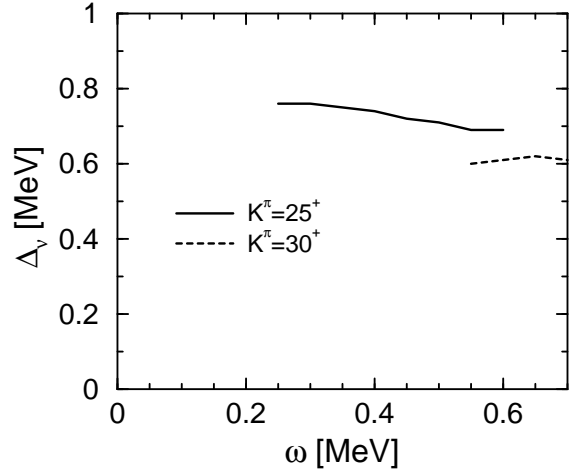,height=6.5cm}} 
        \caption{$\Delta^{PNP}$ as a function of $\omega$ for neutrons 
          in the $K^\pi=25^+$ and $30^+$ 
          bands. Note, in these cases is $\Delta^{HFB}=$0.}  
        \label{fig:deltan25-30} 
\end{figure}

\begin{figure}[hbt] 
           \centerline{\psfig{figure=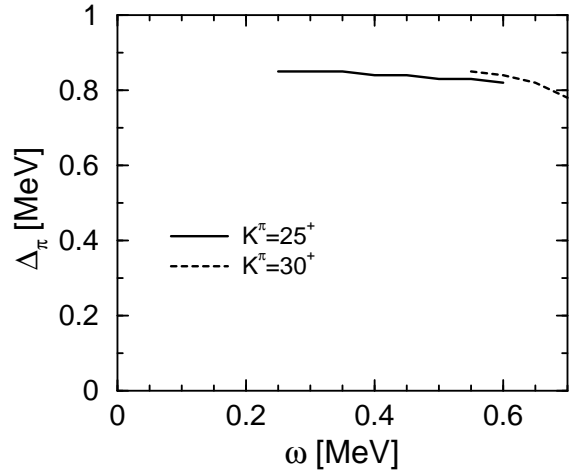,height=6.5cm}} 
        \caption{$\Delta^{PNP}$ as a function of $\omega$ for  
          protons in the $K^\pi=25^+$ and $30^+$ 
          bands. Note, in these cases is $\Delta^{HFB}=$0.}  
        \label{fig:deltap25-30} 
\end{figure}

\unboldmath 
In these eight-quasiparticle configurations the static pair field  
has disappeared ($\Delta^{HFB}=0$). The pair correlations are only of dynamical 
nature. One can expect  the RPA to do a good job in describing 
them because one is sufficiently away from the critical region.  
As mentioned above, 
the PNP calculation leads to a finite value of $\Delta^{PNP}$ which 
is 30$\%$ smaller then the values in the ground state 
(c.f. Fig.~\ref{fig:deltan25-30}~and~\ref{fig:deltap25-30}). There is 
a weak reduction towards higher $\omega$ within each band.   
 
\subsubsection{Routhians} 
\label{RHighK} 
Experimental Routhians are calculated by means of the standard  expressions 
 given e.g. in~\cite{Fr97}.  
In order to remove the steep  decrease of 
$R(\omega)$ a term  $30 MeV^{-1} \omega^2$ is added to all $R$ values, 
which makes the differences   
between the various curves  better visible. 
The calculated ground state energy $R_{g.s.}(\omega=0)$ is set to zero  
for all calculations and the experiment. The PNP and RPA 
calculations give an energy gain relative to the HFB results. Due to 
the blocking effect, this gain is normally smaller in the excited 
states than in the ground state. As a consequence the energy relative 
to the ground state gets larger for RPA and PNP than for HFB 
calculations. This systematic tendency is substantiated by the 
examples to be shown. 
 
\begin{figure}[hbt] 
          \centerline{\psfig {figure=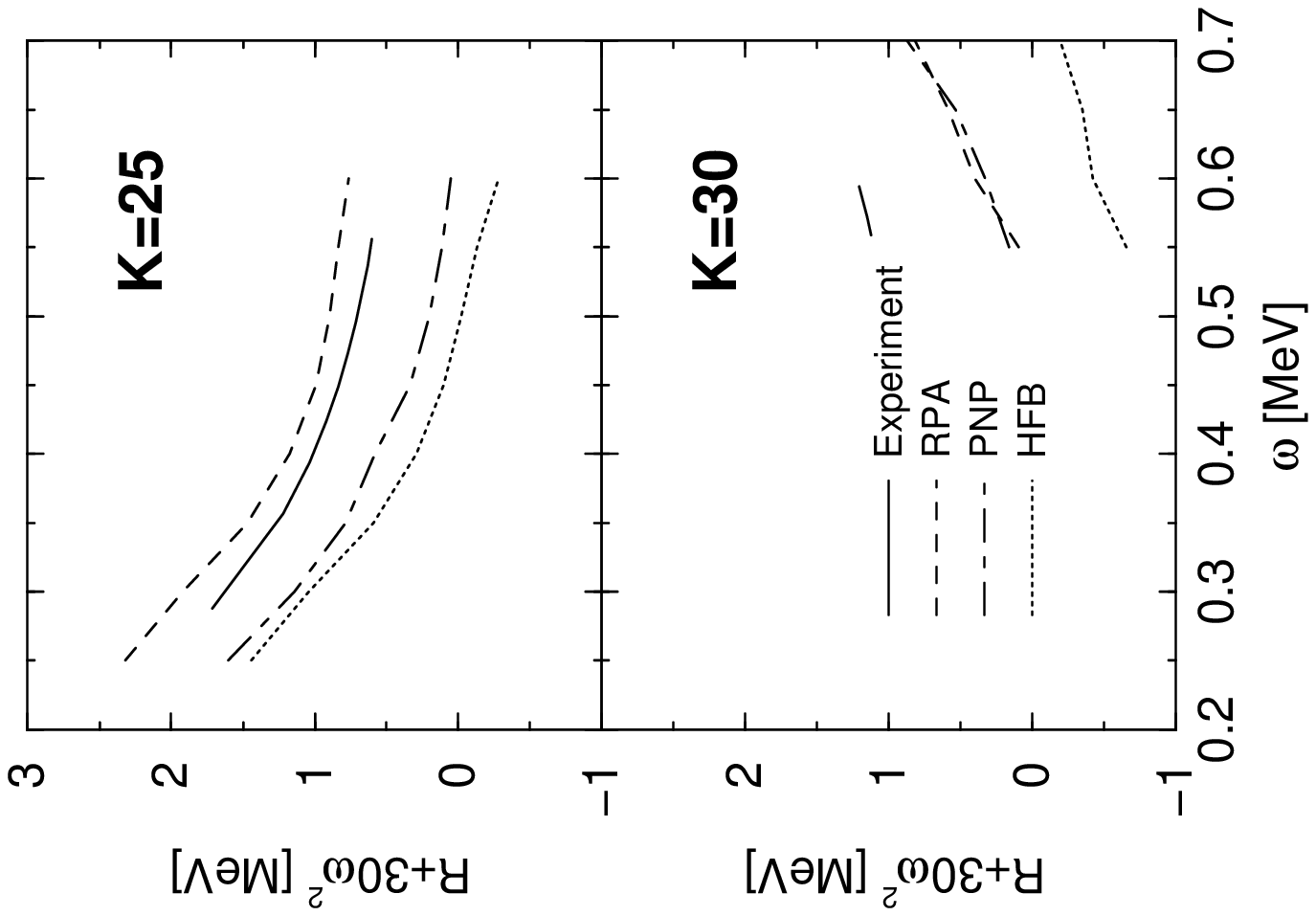,width=8cm,angle=-90}} 
        \caption{The energies in the rotating frame $R(\omega)$ 
          relative to the ground state for 
          the $K^\pi=25^+$ (upper) and $K^\pi=30^+$ (lower) bands. A 
          term of $30 \omega^2$ is added to the $R$ values. The 
          experimental values are taken from~\protect\cite{PW98}.}   
        \label{fig:E'K25-K30} 
\end{figure}

For the $K^\pi=25^+$ band the RPA  
yields a substantially better agreement with experiment than 
PNP (cf. Fig.~\ref{fig:E'K25-K30}). The larger discrepancy for the 
$K^\pi=30^+$ may reflect some  
inaccuracy of single particle levels. Although the HFB calculation 
gives a good estimate of the angular  
momentum~\cite{FN00} it underestimate the relative energy of the 
bands. The rotational frequency of the band head is also quite well 
reproduced by RPA and PNP.  
 
\begin{figure}[hbt] 
    \centerline{\psfig {figure=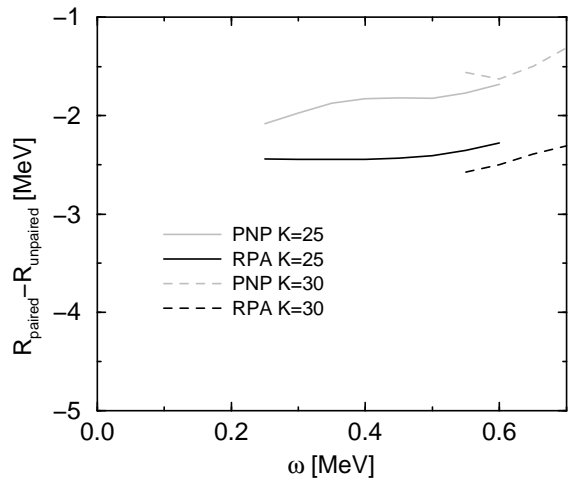,height=6.5cm,angle=-90}} 
  \caption{The total pair correlation energies in the rotating 
    frame $R_{paired}(\omega)-R_{unpaired}$ for the 
          the $K^\pi=25^+$ and $30^+$.}  
        \label{fig:PcorrHigh} 
\end{figure}

In Fig.~\ref{fig:PcorrHigh} the energy differences between the 
paired (PNP or RPA) and unpaired calculation (Hartree-Fock) are 
shown. They characterize the effect of the pair correlation  
onto the energy. 
We can see that the RPA gives a larger pair correlation energy then 
PNP in accordance with~\cite{HB00}. This indicates that the RPA is a 
better approximation than PNP in the $\Delta^{HFB} =0$ regime. One 
notices that the correlation energy is only weakly reduced by rotation. 
 
\subsubsection{Angular momentum} 
\begin{figure}[htb] 
           \centerline{\psfig {figure=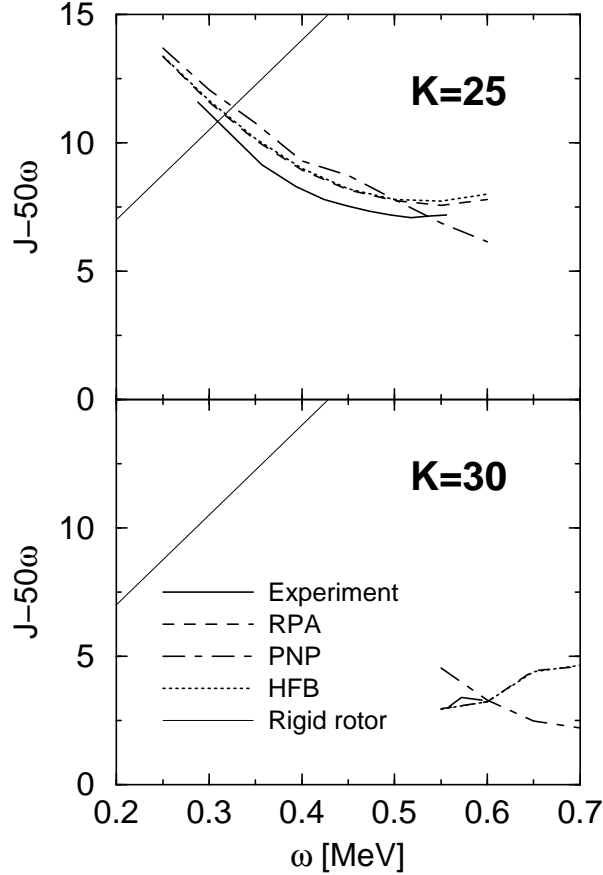,width=8cm,angle=-90}} 
        \caption{Angular momentum $J(\omega)$ for $K^{\pi} = 
          25^+$ and $30^+$. A term of $50 
          \omega$ is subtracted from the $J$ values. The experimental 
          values are taken from~\protect\cite{PW98}. The RPA and HFB 
          results practically coincide in this picture.} 
        \label{JhighK} 
\end{figure}

The investigation \cite{FN00} demonstrated that the angular momentum 
and the dynamical moment of inertia of $^{178}W$ can be understood by 
assuming that the nucleons move in a rotating mean field with no 
pairing. The strong reduction of the moment of inertia as compared to 
the rigid body\footnote{This limit is  
referred to the rotational spectrum $E(I)=A_{rig}I(I+1)-K^2$.} 
value is due to the nuclear shell structure (cf.~\cite{FN00} and 
section~\ref{allK}).  As seen in Fig.~\ref{JhighK}, the inclusion of dynamic pair correlations does 
not change this result. Note 
that a linear term has been subtracted  so that 
the differences between the curves are considerably magnified.
All approaches 
reproduce the experimental dynamical moment of inertia (${\cal 
  J}^{(2)} = \frac{dJ}{d\omega}$) which is substantially below the 
rigid body value.  
The differences of the function $J(\omega)$ between paired and the 
unpaired calculations are only marginal. The dynamical pair correlations do not change the 
angular momentum very much, but they do increase the energy 
difference between bands with a different number of blocked 
quasiparticles.  
  
\subsubsection{Branching ratios} 
\label{branching} 
In the TAC model, the $B(M1)$ and $B(E2)$ values are
calculated by means of the expressions  
\begin{equation} 
        B(M1)=\frac{3}{8\pi} \left[\sin\vartheta \left(J_{3\pi}+2.35 
            S_{3\pi}-2.24 S_{3\nu}\right) -\cos\vartheta 
          \left(J_{1\pi}+2.35 S_{1\pi}-2.24 S_{1\nu}\right)\right]^2  
        \label{BM1} 
\end{equation} 
and 
\begin{equation} 
        B(E2)=\frac{15}{128\pi}\left(\sin\vartheta\right)^4 Q_{0}^{2} 
        \label{BE2} 
\end{equation} 
where $J$, $S$ and $Q_{0}$ are the expectation values of the angular 
momentum, the spin and the quadrupole moment, respectively, as 
calculated with the TAC states. It is common practice to present the 
experimental branching ratios in the form 
$\left|\frac{(g_K-g_R)}{Q_0}(\omega)\right|$, which is obtained 
assuming that the strong coupling limit~\cite{BM69} is valid. We 
choose to display the calculated branching ratios (which of course do 
not rely on the strong coupling assumption) in the same way. The 
theoretical ratios are obtained as  
\begin{equation} 
        \left|\frac{(g_K-g_R)}{Q_0}(\omega)\right| = 
        \sqrt{\frac{5}{16}} \sqrt{\frac{1}{K^2}-\frac{1}{J^2}} 
        \sqrt{\frac{B(M1)}{B(E2)}}  
        \label{gK-gR} 
\end{equation} 
where $K$, the value of the angular momentum at the band head, is kept 
constant and $J$ is the calculated value of the angular 
momentum. $Q_0$ is chosen as in~\cite{PW98}.   
 
\begin{figure}[hbt] 
           \centerline{\psfig {figure=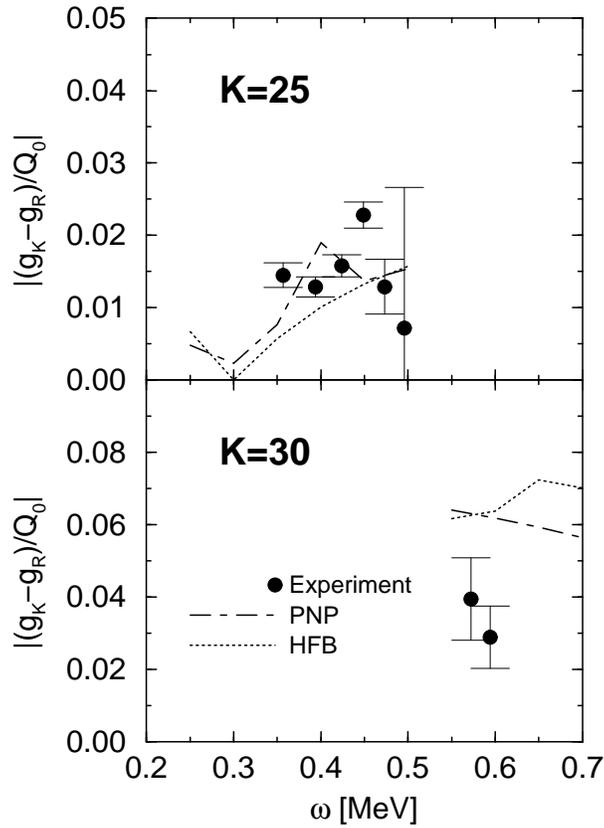,width=8cm,angle=-90}} 
        \caption{The ratio $|\frac{(g_K-g_R)}{Q_0}(\omega)|$ for the 
          $K^\pi=25^+$ and $30^+$ bands. The experimental values are taken 
          from~\protect\cite{PW98}. Observe the different scales.}  
        \label{ratiohighK} 
\end{figure} 
The theoretical and experimental ratios are given in 
Fig.~\ref{ratiohighK}. Since the ratio depends  
strongly on the orientation of the proton and neutron angular  
momentum the good agreement confirms our calculated geometry. The 
experimental errors of the branching ratios are too  
large to discriminate between the PNP and HFB calculation.  
 
\boldmath 
\subsection{$K^\pi=7^-$, $15^+$ and $22^-$}\label{71522}
\unboldmath 
The $K^\pi=7^-$, $15^+$ and $22^-$ bands are analyzed and presented in 
the same way as the $K^\pi=25^+$ and $30^+$ bands.  
These bands, which correspond to the excitation of two to six quasiparticles 
from the $K=0$ ground configuration, have a reduced but non-zero     
static HFB pair field at the band head.  
The PNP is expected to be 
stable while the RPA can run into problems since one is close   
to the transition $\Delta^{HFB} \rightarrow 0$. 
 
\begin{figure}[hbt] 
           \centerline{\psfig {figure=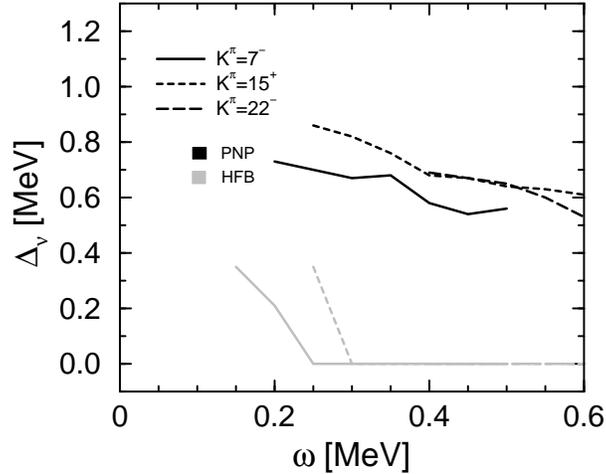,height=6.5cm}} 
   \caption{$\Delta^{HFB}$(grey) and $\Delta^{PNP}$(black) as a 
     function of $\omega$ for neutrons in the $K^\pi=7^-$, 
     $15^+$ and $22^-$ bands.}  
        \label{fig:deltan7-22} 
\end{figure} 
\begin{figure}[hbt] 
           \centerline{\psfig {figure=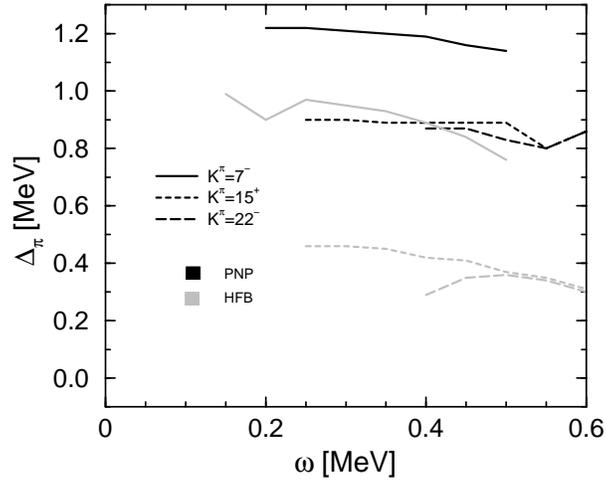,height=6.5cm,angle=-90}} 
   \caption{$\Delta^{HFB}$(grey) and $\Delta^{PNP}$(black) as a 
     function of $\omega$ for protons in the $K^\pi=7^-$, 
     $15^+$ and $22^-$ bands.}  
        \label{fig:deltap7-22} 
\end{figure} 
Figure~\ref{fig:deltan7-22} and~\ref{fig:deltap7-22} show the 
calculated values of $\Delta$. The values of 
$\Delta^{PNP}$ are similar  to the ones  for the  
$K^\pi=25^+$ and $30^+$ bands in Figs.~\ref{fig:deltan25-30} 
and~\ref{fig:deltap25-30}, except $K^\pi=7^-$, for 
which the protons remain in the ground configuration.  
For the two quasi neutron bands $K^{\pi}=7^{-}$ and $15^+$ the gap 
$\Delta^{HFB}_{\nu}$ is reduced strongly already at the band head and 
becomes zero at $\omega \approx 0.3$ (MeV). 
These values indicate that the pair filed is nearly instable, which 
causes problems to be be discussed in 
section~\ref{AmLowK}. Though $\Delta^{HFB}_{\pi}$ is strongly reduced in 
the two quasi proton configurations $K^\pi=15^+$ and $22^-$ it changes  only 
weakly with $\omega$. 
 
\subsubsection{Routhians} 
\label{RLowK} 
Both the PNP and RPA calculations fairly well reproduce  
the Routhians at high $\omega$. There, again the tendency is seen 
that the dynamical pair correlations enlarge the energy distance 
between the bands. The exception is the $K^\pi=15^+$ for $\omega>0.38$ MeV. 
This discrepancy will be discussed in section~\ref{AmLowK}. 
\begin{figure}[htb] 
           \centerline{\psfig {figure=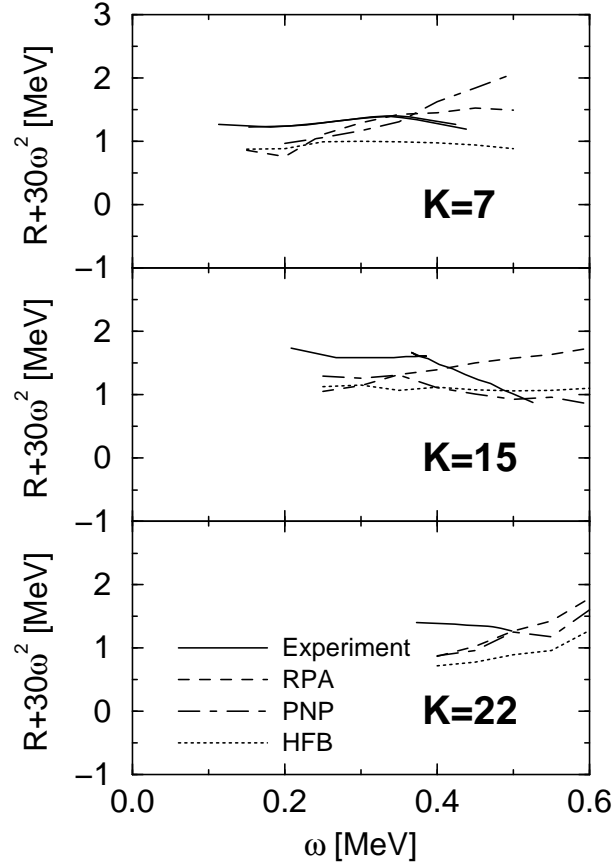,width=8cm,angle=-90}} 
        \caption{The energies in the rotating frame $R(\omega)$ 
          relative to the ground state for 
          the $K^{\pi} = 7^{-}$ (upper), $15^+$ (middle) and $22^-$ 
          (lower) bands. A term of  
          $30 \omega^2$ is added to the $R$  
          values. The experimental values are taken 
          from~\protect\cite{PW98}.}  
        \label{E'K7-K22} 
\end{figure} 
 
\begin{figure}[hbt] 
    \centerline{\psfig {figure=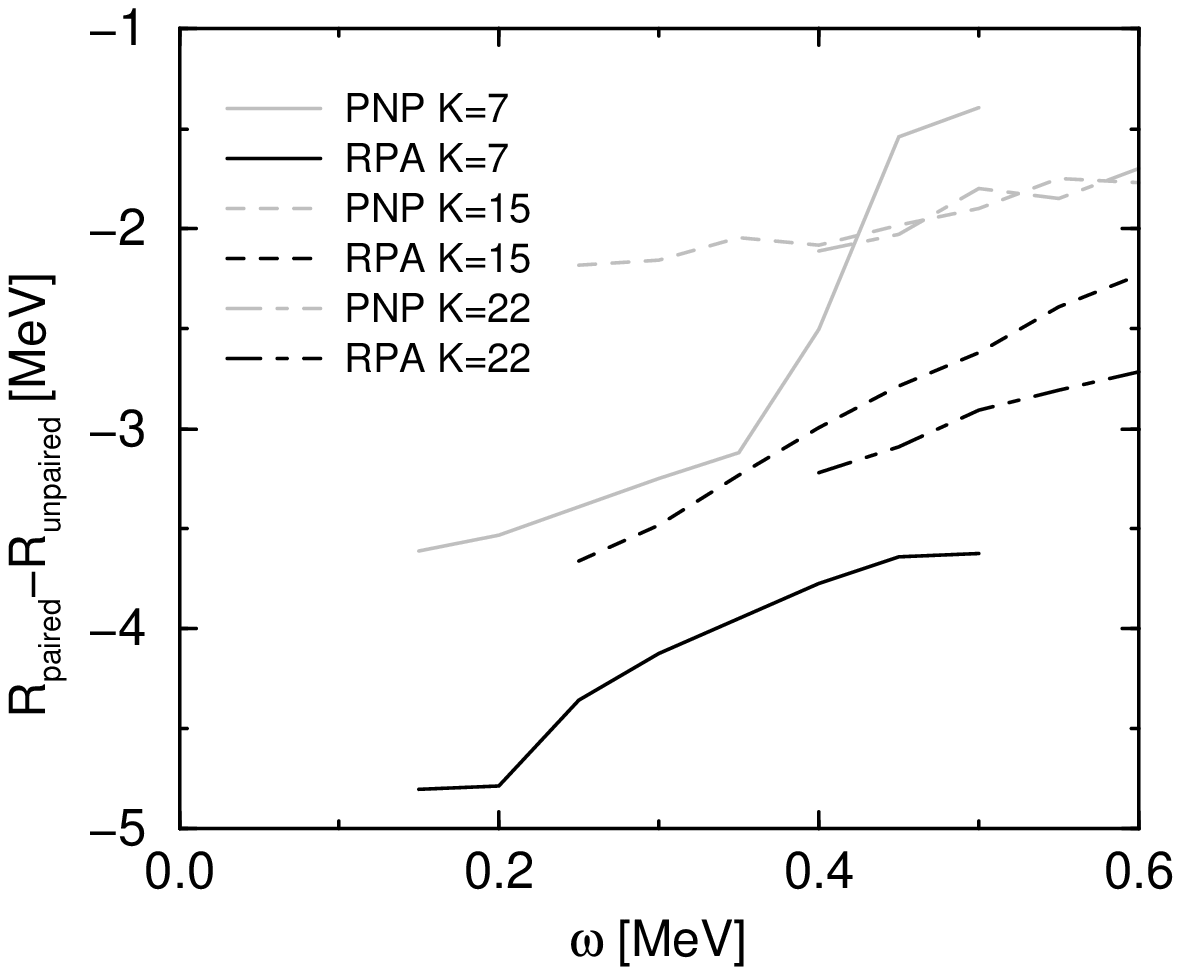,height=6.5cm,angle=-90}} 
  \caption{The total pair correlation energies in the rotating 
    frame $R_{paired}(\omega)-R_{unpaired}$ for the 
          the $K^\pi=7^-$, $15^+$ and $22^-$.}  
        \label{fig:PcorrLow} 
\end{figure}

Figure~\ref{fig:PcorrLow} shows 
that the pair correlation energies  
for these lower $K$ values are larger and depend stronger on the 
rotational velocity than the ones in Fig.~\ref{fig:PcorrHigh}. The HFB + RPA 
calculations cannot reproduce the backbend in the $K^\pi=7^-$ and 
$15^+$ bands since $\Delta^{HFB}_\nu =0$ for $\omega > 0.25$ MeV.   
It  is seen as the kink in Fig.~\ref{E'K7-K22} and the  
upbend of the PNP calculation in Fig.~\ref{fig:PcorrLow}. 
 The erratic behavior of the PNP pairing energy at large $\omega$ 
can be understood by the 
fact that the energy surface has a shallow minimum and is therefore 
easily disturbed by structure effects caused by quasiparticle 
crossings. Most of these crossings do not exist in the HFB + RPA 
calculations because $\Delta^{HFB}_\nu =0$. As a consequence, $R$ 
depends more smoothly on $\omega$, as the experiment does. We consider  
the fluctuations of $R$  as an error of the  PNP approach,  
which mimics the dynamical correlations by a static $\Delta^{PNP}$. 
 
\subsubsection{Angular momentum} 
\label{AmLowK} 
\begin{figure}[htb] 
           \centerline{\psfig {figure=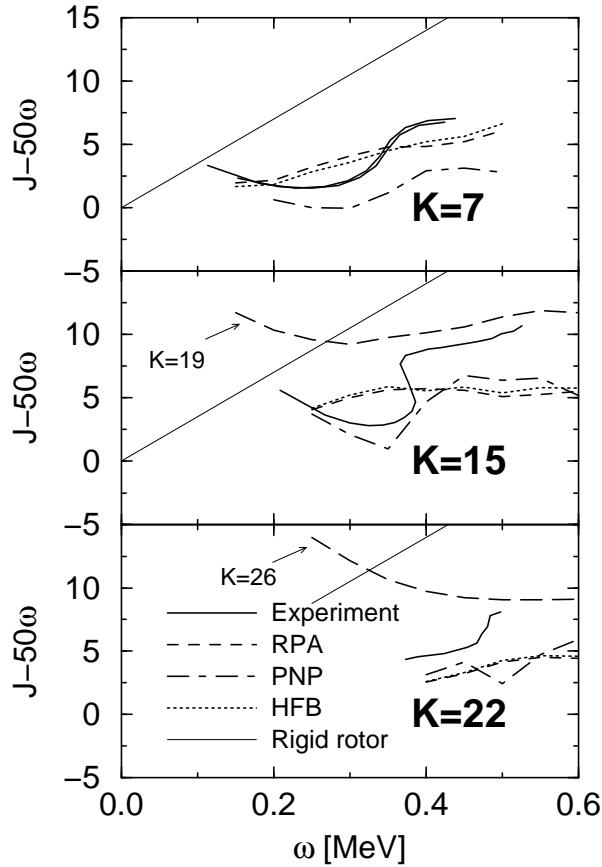,width=8cm,angle=-90}} 
        \caption{Angular momentum $J(\omega)$ for $K^{\pi} = 
          7^{-}$, $15^+$ and $22^-$. The RPA solution for the 
          $K^\pi=19^+$ (middle) and $K^\pi=26^-$ (lower) are 
          added. A factor of 
          $50 \omega$ is subtracted from the $J$  
          values. The experimental values are taken   
          from~\protect\cite{PW98}. The RPA and HFB 
          results practically coincide in this picture.}  
        \label{JlowK} 
\end{figure}

As seen in  Fig.~\ref{JlowK}, the calculations give reasonable agreement 
with the experiment. The neutron backbends in 
the $K^\pi=7^-$ and  
$15^+$ are not reproduced in the HFB + RPA calculation because the static 
pair gap collapses already before the backbend. This feature could be 
corrected by using a somewhat larger pairing strength $G_\nu$. The PNP  
calculation gives approximately the right value of  $\omega_{c}$ for 
the neutron backbend but gives a slightly too low value of $J(\omega)$ 
for the $K^{\pi} = 7^{-}$ band.  
 
In the $K^\pi=15^+$ band, all the calculations underestimate the 
angular momentum by $\sim 5 \hbar$ after the backbend at  
$\omega \sim 0.38$  
MeV. We assign this increase to a configuration change caused by  
a crossing between a   
$h_{9/2}$ proton orbital and a $h_{11/2}$ proton orbital. As a result  
 a $K^\pi=19^+$ band continues the  $K^\pi=15^+$ 
band. This crossing occurs at the same $\omega$ as  
the neutron crossing causing the normal backbend. The same proton 
crossing occurs in the $K^\pi=22^-$ band at $\omega \sim 0.48$ MeV 
which continues as a $K^\pi=26^-$ band. It is seen in  
Fig.~\ref{JlowK}  as an upbend. 
The configurations of the $K^\pi=19^+$ and $26^-$ bands are 
listed in table~\ref{config2}. The calculations for the $K^\pi=7^-$ 
band after the backbend also  
gives a bit too low angular momentum. It cannot be  explained by the 
 proton crossing discussed. This may indicate that the missing 
angular momentum in the $K^\pi=15^+$ and $22^-$ bands has another 
origin than suggested above.  
 \begin{table}[htb] 
\begin{center} 
\begin{tabular}{crl} 
$K^{\pi}$ & Neutron configuration       & Proton configuration\\ \hline 
$19^{+}$  & $\nu\{7/2^{+},7/2^{-}\}$ & 
$\pi\{1/2^-,7/2^-,7/2^{+},9/2^{-}\}$ \\  
$26^{-}$  & $\nu\{5/2^{-},7/2^{+},7/2^{-},9/2^{+}\}$ & 
$\pi\{1/2^-,7/2^-,7/2^{+},9/2^{-}\}$ \\   
\end{tabular} 
\caption{The quasiparticle configurations used in the article are 
  taken from~\protect\cite{PW98}. (Neutrons: 
  $5/2^{-}[512],7/2^{+}[633],7/2^{-}[514],9/2^{+}[624]$ Protons: 
  $1/2^{-}[541],7/2^{-}[523],7/2^{+}[404],9/2^{-}[514]$)}  
\label{config2} 
\end{center} 
\end{table}

\subsubsection{Branching ratios} 
\begin{figure}[hbt] 
           \centerline{\psfig {figure=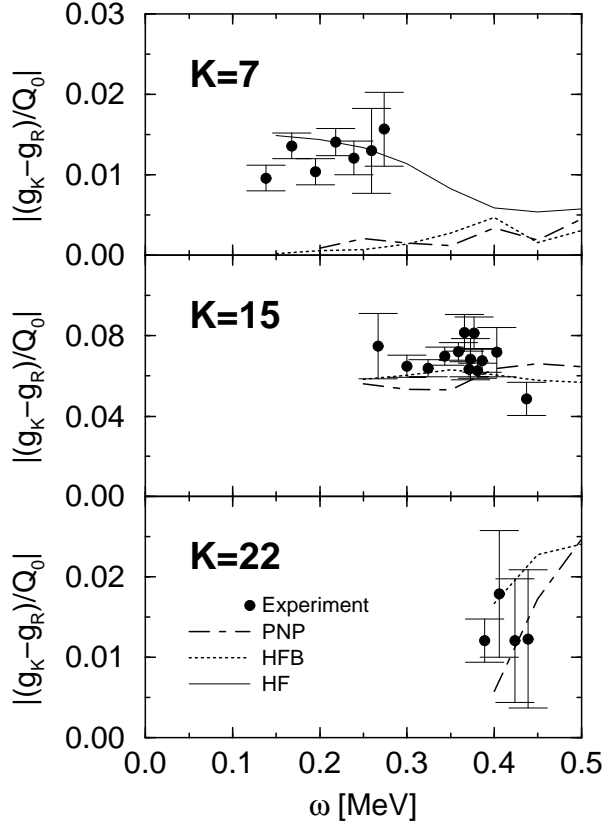,width=8cm,angle=-90}} 
         \caption{The ratio $|\frac{(g_K-g_R)}{Q_0}(\omega)|$ for the 
          $K^\pi=7^-$ (upper), $15^+$ (middle) and $22^-$ (lower) 
          bands. The thin full line  
          (HF) displays the case $\Delta=0$  
          for comparison. The experimental values are taken 
          from~\protect\cite{PW98}. Observe the different scales.}  
        \label{ratiolowK} 
\end{figure}

The good agreement with experiment in Fig.~\ref{ratiolowK} for the 
$K^\pi=15^+$ and $22^-$ confirms the geometry of our calculations.  
Again, the large experimental errors of the branching ratios do not allow to 
distinguish between the PNP and HFB calculations.   
 
In the case of the $K^{\pi} = 7^{-}$ band, the 
 calculation with $\Delta =0$  reproduces the branching ratios better.  
We consider this as accidental, because the presence of the upbend  
seen in the $I(\omega)$ curve of Fig.~\ref{JlowK} is a clear indication  
of a finite pairing gap for the neutrons. The protons are  
definitely in their ground configuration with a non-zero gap. 
The $K^{\pi} = 7^{-}$ band has small $B(M1)$ values compared to  
other $K$-bands. This is expected because  
the normally dominating  
proton contributions to the $B(M1)$ value (\ref{BM1})  
are small for the ground configuration and very sensitive to the size
 of $\Delta_\pi$,  
It is largely   canceled by the negative neutron spin contribution.  
Hence, the resulting ratio is not reliable.

\boldmath 
\subsection{COMPARISON OF ALL $K$-BANDS} 
\begin{figure}[hbt] 
    \centerline{\psfig {figure=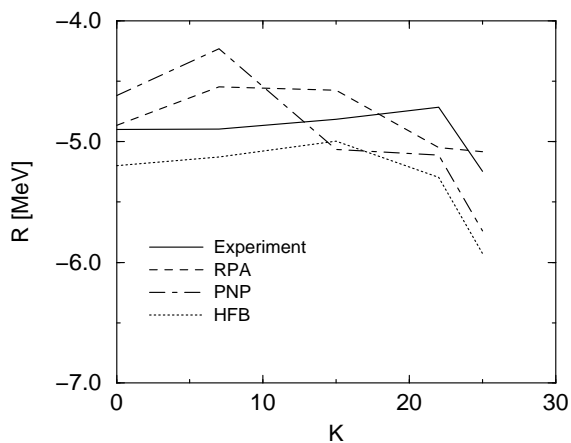,width=7.5cm,angle=-90}} 
  \caption{The energies in the rotating frame relative to the ground 
    state as a function of  
    $K$ for $\omega=0.45$. The $K$-values plotted are $0$ 
    (s-band)$, 7, 15, 22$ and $25$. The experimental values are 
    taken from~\protect\cite{PW98}.}  
  \label{RK} 
\end{figure}

\begin{figure}[hbt] 
    \centerline{\psfig {figure=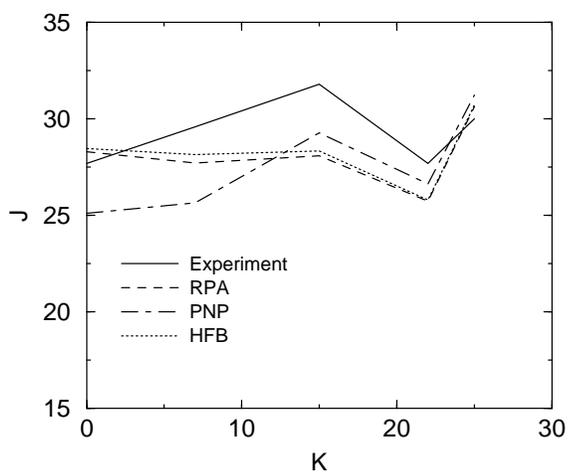,width=7.5cm,angle=-90}} 
  \caption{The angular momentum as a function of 
    $K$ for $\omega=0.45$. The $K$-values plotted are $0$(s-band)$, 7, 
    15, 22$ and $25$. The experimental values are  
    taken from~\protect\cite{PW98}.}  
  \label{IK} 
\end{figure}

\begin{figure}[hbt] 
    \centerline{\psfig {figure=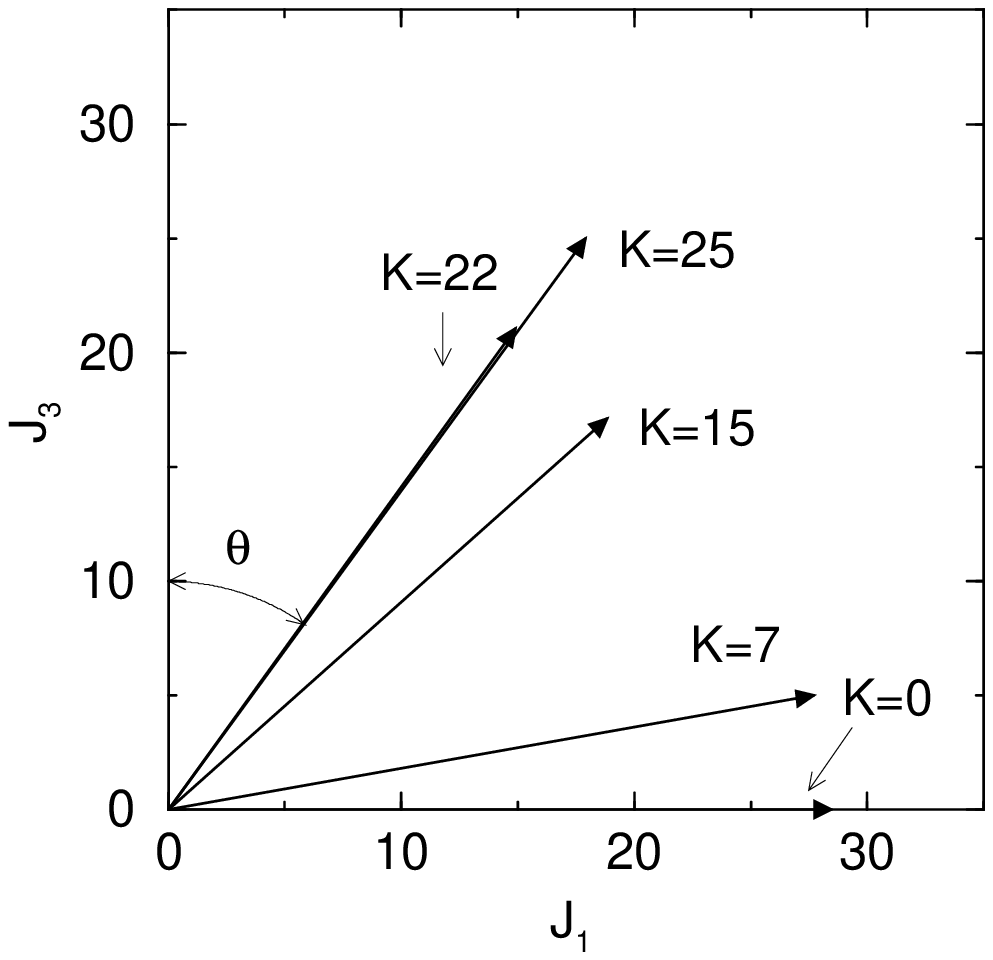,width=7.5cm,angle=-90}} 
  \caption{The angular momentum component along the 1- and 3-axis for 
    different $K$ at $\omega=0.45$ as calculated with HFB. The 
    $K$-values plotted are $0$(s-band)$, 7, 15, 22$ and $25$.}  
  \label{TK} 
\end{figure}

\unboldmath 
\label{allK} 
In Figs.~\ref{RK}--\ref{TK} we compare the Routhians and the angular 
momenta of  
different $K$-bands at a fixed frequency $\omega=0.45$ MeV. The $K=30$ band is 
left out from the plots since it starts at a higher $\omega$. The 
bands with different orientation of $\vec{J}$ have approximately the 
same energy $R$ in the rotating frame. Their energies in the laboratory 
frame for a given value of $J$ do not differ much as well. That is so, because 
 the energy needed for generating angular momentum along the two principal 
axes 1 and 3 is nearly the same. This feature, which is reflected by 
many bands with different orientation being close to the yrast line, 
is characteristic for the ($N$,$Z$) region where $^{178}W$ is  
situated. 
 
 
\begin{figure}[htbp] 
    \centerline{\psfig {figure=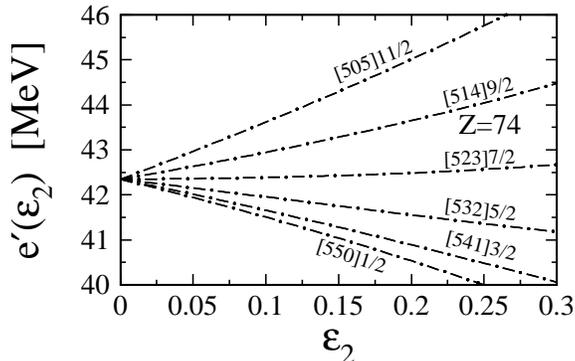,width=7.5cm,angle=-90}} 
  \caption{The negative parity states in the 50--82 protons 
    shell. The larger gaps between the levels in the upper part of 
    the shell promote deformation alignment of angular momentum. The  
    levels are labeled with the asymptotic quantum numbers $[ N n_z  
    m_l ] \Omega$.} 
  \label{fig:pN5} 
\end{figure} 
In the upper and middle part of the 50--82 proton and 
82--124 neutron shells  
there are many orbitals with higher $\Omega$ values  
that strongly couple to the prolate deformed potential 
whereas the low-$\Omega$ orbitals are occupied. 
 Figure~\ref{fig:pN5} shows the $h_{11/2}$ orbitals as an example.  
For the high-$\Omega$ orbitals 
the Coriolis coupling between states with $\Omega$ 
and $\Omega +1$, which generate a contribution to $J_1$, is weak, 
because the energy distance is large and the coupling matrix-element 
($\sim \sqrt{j^2-\Omega^2}$) is small. This is very different from the  
nuclei situated at the bottom of the shell.  
For the low-$\Omega$ orbitals  the energy distance 
is small and the coupling matrix element large. Hence it is easy to 
generate a contribution to $J_1$.  
Thus, the strong coupling  suppresses
the angular momentum alignment along the collective 1-axis,    
which causes the reduction of the moment of inertia  
relative to the rigid body value.  
 
On the other hand, it is energetically  favorable to generate  
a contribution to $J_3$, because there are many orbitals with a large 
$\left| \Omega \right|$ near the Fermi surface and it does not cost 
much energy to generate a particle-hole excitation with a large 
$K=\Omega_p-(-\Omega_h)$. At the bottom of the shell these excitations
cost much more energy, because the orbitals with large $\left| \Omega 
\right|$ are far away. The net result is that at the beginning of the 
rare earth region the moment of inertia is larger than the rigid body 
value and generating angular momenta along the 3 axis is 
unfavorable. In the {\em upper part} of the region, the moment of 
inertia is smaller than the rigid body value and generating angular 
momenta along the 1 and 3 axis is equally favorable (see~\cite{FN00} 
and~\cite{PF75}). 
The inclusion of pairing 
does not change this conclusion.  
 
Figure~\ref{TK} shows  
the $J_1$ and $J_3$ components  
for different $K$-bands calculated with HFB at a  typical frequency 
$\omega=0.45$. 
Within $\pm 2 \hbar$, the component  $J_3$ is equal to $K$ which means
that all bands are not far from  the strong coupling regime.
The largest deviation is found in the $K=15$ band 
where $J_3\approx 17$ at $\omega=0.45$. This is an effect of the 
configuration mixing in the neutron system, which is induced by 
quasi crossings in a region of high quasiparticle level density. 
The total spin  $J$ is almost independent of $K$,
in accordance with the discussion above.
Generating the components $J_1$ and  $J_3$ of angular momentum 
costs a comparable amount of energy.
The slightly different behavior of the $K=25$ band is due to  
extra rotational alignment from  
the aligning $h_{9/2}$ proton orbital, see discussion in~\cite{FN00}.

Summarizing our calculations concerning an improved treatment of pairing 
we find that i) the conclusions of earlier work~\cite{FN00} can be confirmed 
 and ii) the angular momentum  
$J(\omega)$ is insensitive to the pair correlations in the frequency 
region after the first band crossing.  
This is at variance with the results of \cite{SG89,BD86} for low-K 
bands, where a reduction of several units of angular momentum by 
the dynamical pair correlations was found. The reason is
that the single particle  
orbits near the Fermi surface, which contribute most to the collective 
angular momentum, are blocked for the pair correlation by generating 
the high K. The ratio $|\frac{(g_K-g_R)}{Q_0}(\omega)|$ is also 
reasonably well reproduced which support the calculated tilted geometry 
of our solutions.  
        
\section{CONCLUSIONS} 
We included dynamical pair correlations into the tilted axis cranking 
approach in two different ways: The pairing-RPA  
method, which allows for harmonic vibrations on top of the HFB mean field, 
and the particle  
number projection which describes the dynamical correlations by an  
increased static pair gap in conjunction with projection onto good particle  
number. We studied the high $K$-bands in 
$^{178}W$  by means of both methods. As known from   
investigation of the band head energies~\cite{JB95}, 
the HFB approach tends to underestimate the energetic separation between  
the bands. Inclusion of the dynamic correlations improves the
energy of bands with zero static pair correlations relative to each other
and to the bands with pronounced static pair correlations.
 When the static pair field is substantial we could not find a 
systematic improvement of the relative energy due to the dynamical pair
correlation. This seems to be in contrast to the result for $K=0$ bands 
in~\cite{SG89,BD86}.  
 
The pronounced reduction of the moment of inertia of high-$K$ bands  
relative to the rigid body  
value, seen in the experiment, is not due to the pair correlations. It 
is caused by orbitals close to the Fermi surface which in the 
mid and upper part of its shell is disfavoring the generation of 
angular momenta perpendicular to the symmetry axis. 
It is typical for this region, that the generation of  
angular momentum along the symmetry axis by particle-hole excitation 
is equally favorable as compared to the rotational alignment i.e.  
collective rotation along the perpendicular axis (see~\cite{FN00} 
and~\cite{PF75}). The inclusion pair correlations, both static and  dynamic,
  does not 
significantly affect this feature of high-$K$ bands, which is
in stark contrast to low-$K$ bands.  
 
In bands where the static pair gap of the HFB treatment has collapsed to zero  
the RPA method is simpler to use than PNP because there is no need to  
minimize the Routhian with respect to the $\Delta$ and $\lambda$ 
parameters. The pair gap introduced in these PNP calculations 
causes irregularities in the energy and angular momentum which are 
caused by quasiparticle band crossings.  
The RPA method, which does not have these crossings, describes the 
experimental results better. Thus, we consider the irregularities 
appearing in PNP as an artifact of the approximation and RPA superior 
in this region. On the other hand,  
when the HFB mean field pair gap is reduced but not yet zero  
the RPA method has problems because of its deficiencies near the 
transition to zero pairing. In these cases it turns out to be 
important to check the  
quality of the quasiboson approximation when calculating the RPA 
contribution to the angular momentum.  
The PNP method is  more stable in these cases. 
 
We gratefully acknowledge  valuable discussions with R. G. Nazmitdinov. 
Work is supported by the DOE grant DE-FG02-95ER40934.

\appendix 
\section{THE ROLE OF THE EXCHANGE TERM IN PARTICLE NUMBER PROJECTION} 
\label{PNP} 
The HFB pairing energy is usually calculated in Hartree approximation, 
i.e. neglecting the exchange term by factorizing the pairing matrix 
element $<P^\dagger P>\, \approx\, <P^\dagger><P>\,$. This 
approximation is justified for the calculation of the pairing energy 
contribution without PNP. However, it was recently suggested 
~\cite{Do98} that the neglect of exchange terms in performing PNP can 
lead to dangerous poles in the resulting total Routhian surface (TRS) 
 Such an unphysical behavior of the PES was indeed found in our 
calculations and it was traced back to the above mentioned 
factorization. For the sake of completeness we sketch the 
argumentation presented in more detail in ~\cite{Do98}.  
   
Using in the construction of the PNP state $|N>$ 
eq.~(\ref{projection}) and  the canonical 
(BCS-like) form~\cite{RS80} of the HFB state $|>$, the number 
projected pairing energy $E_{pair} = -G <N|P^\dagger P|N>$ can be 
written as the sum  
\begin{equation} 
E_{pair} = E_{pair}^{Direct} +   E_{pair}^{Exchange} , 
\end{equation} 
where the direct term  
\begin{equation} 
  E_{pair}^{Direct} = - G <N|N>^{-1} \int\!\! d\varphi\left[ \sum_{k>0}  
    \frac{2 P_{k\bar{k}}  v_k u_k e^{i 
        \varphi}}{\left(u_{k}^{2}+v_{k}^{2} e^{2i 
          \varphi}\right)}\right]^2 <\varphi=0|\varphi>.  
  \label{direct} 
\end{equation} 
and the usually neglected exchange term 
\begin{equation} 
 E_{pair}^{Exchange} = -G  <N|N>^{-1} 
\int\!\! d\varphi \sum_{km} \frac{2 P_{km}^2 v_{k}^{2} v_{m}^{2} 
  e^{4i\varphi}}{\left(u_{k}^{2}+v_{k}^{2} e^{2i 
      \varphi}\right)\left(u_{m}^{2}+v_{m}^{2} e^{2i \varphi}\right)} 
<\varphi=0|\varphi> .   
        \label{exchange} 
\end{equation} 
These expressions contain the canonical BCS-amplitudes $u_k, v_k$ of 
the quasiparticle state and the matrix elements $ P_{kk'}$ of the pair 
operator in the canonical basis. The bracket   
\begin{equation} 
<\varphi=0\,|\,\varphi> \, \equiv \,\Pi_{k>0}\, (u_k^2 + v_k^2 e^{2i 
  \varphi}).  
        \label{overlap} 
\end{equation} 
is the overlap function between gauge rotated quasiparticle 
states.  
 
One may encounter a zero denominator $\left(u_{k}^{2}+v_{k}^{2} e^{2i 
    \varphi}\right)$ in both energy terms 
(\ref{direct},\ref{exchange}) e.g. for $u_k = v_k$ and $\varphi = 
\frac{\pi}{2}$. This is because the double zero in the denominator 
cannot be canceled by the corresponding single zero in the overlap 
$<\varphi=0|\varphi>$, eq.~(\ref{overlap}). However, when summing up 
the two contributions~(\ref{direct},\ref{exchange}) to the full 
pairing energy $E_{pair}$ such unphysical poles do exactly cancel.  
 
\begin{figure}[hbt] 
           \centerline{\psfig {figure=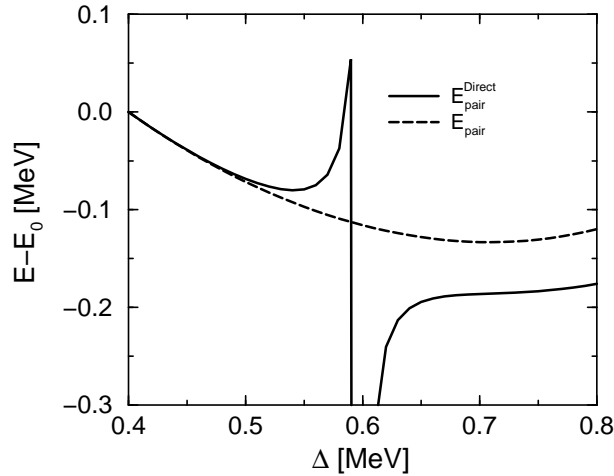,height=6.5cm,angle=-90}} 
        \caption{The energies $E_{pair}$ and the 
          $E_{pair}^{Direct}$ calculated for $\omega=0.3$ in the 
          $K^{\pi}=7^-$ band with $\lambda_N=49.54$. Both curves are 
          normalized to zero at $\Delta =0.4$.}  
        \label{Eproj} 
\end{figure}  
Our calculations confirm the conclusion that a reliable calculation of 
the TRS with PNP has to be done with the full expression. In 
Fig.~\ref{Eproj} the full neutron energy  is shown as a function of 
the neutron gap ($\Delta$) and it is compared to the one where only 
the direct pairing energy term~(\ref{direct}) is taken into 
account. The full energy has the expected parabolic shape with the 
minimum whereas the curve of the direct term alone displays an 
unphysical pole around $\Delta \approx 0.6$ MeV. 
\begin{figure}[htb] 
           \centerline{\psfig {figure=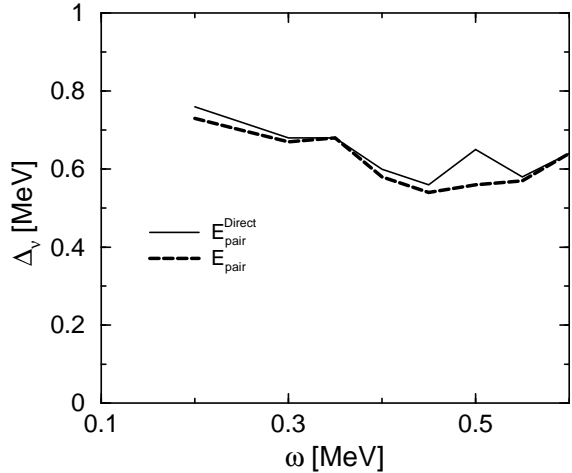,height=6.5cm,angle=-90}} 
        \caption{$\Delta$ determined from $E_{pair}^{direct}$ and 
          $E_{pair}$, respectively,  for the $K^{\pi}=7^-$ band, 
          see also 
          Fig.~\ref{fig:deltan25-30}~and~\ref{fig:deltan7-22}. There is 
          only a small difference except at $\omega=0.5$ where there 
          is a pole in the energy function $E_{pair}^{direct}$ , see 
          Fig.~\ref{Eproj}. These values of $\Delta$ were found 
          keeping all other parameters constant at the self-consistent 
          values found using the full projection.}  
        \label{delta-pnp} 
\end{figure} 
\begin{figure}[hbt] 
           \centerline{\psfig {figure=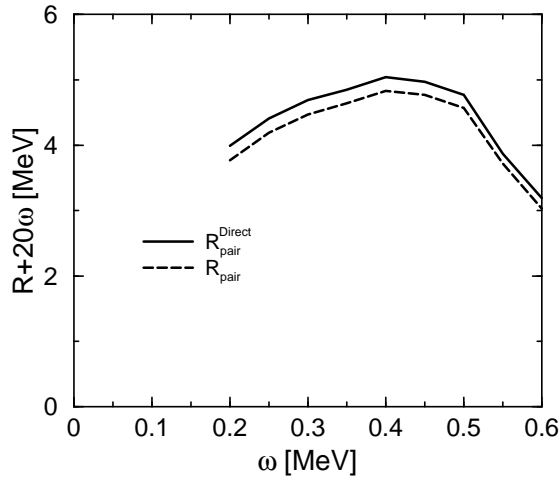,height=6.5cm,angle=-90}} 
        \caption{Routhians $R$ calculated with $E_{pair}^{direct}$ and 
          $E_{pair}$, respectively, for the $K^{\pi}=7^-$ band, 
          see also Fig.~\ref{E'K7-K22}. There is only a marginal 
          difference between both Routhians where all other 
          parameters are kept constant at the self-consistent values found 
          using the full projection.}  
        \label{E'-pnp} 
\end{figure} 
 Such a strange behavior does not happen often and 
usually there is only a minor difference between the extracted 
$\Delta$-value at the minimum for the full and direct energy 
(cf. Fig.~\ref{delta-pnp}). Observables like the energy are not strongly 
affected by the exchange term except close to poles 
(cf. Fig.~\ref{E'-pnp}). With the full pairing energy, we obtain a 
minimum for a slightly  
different $\Delta$ and, therefore, a different $G$ is needed to match 
the experimental $\Delta$-values. We found that a $G_\pi$ of $0.121$ 
MeV instead of $0.119$ MeV should be used for the protons while the 
$G_\nu$ did not change when using the full expression for the pairing 
energy.  
 
The probability of accidentally hitting a pole is not large but it 
happened a couple of times in our calculations. The tail 
(cf. Fig.~\ref{Eproj}) of a pole can 
also affect the results and this is, of course,  much harder to detect. The 
energy surface calculated with only the direct term jumps when 
passing through the pole. This is because a pole has gone in (or out) 
to the area in the complex plane around which we are integrating. The 
pole would turn into a step function if it was possible to perform the 
integration exactly. In order to avoid such unphysical  one should 
generally apply the full expression of the PNP pairing energy.   
 
\end{document}